\documentclass[aps,twocolumn,english,reprint,longbibliography, superscriptaddress, breaklinks=true, showkeys, showpacs=false, nofootinbib]{revtex4}
\usepackage[T1]{fontenc}
\usepackage[utf8]{inputenc}
\usepackage{color}
\usepackage{babel}
\usepackage{amsmath}
\usepackage{amssymb}
\usepackage{subfigure}
\usepackage{graphicx}
\usepackage{physics}
\usepackage{longtable}
\usepackage{amsthm}
\usepackage{tabularray}
\usepackage{tabularx}
\usepackage{array}
\usepackage{float}
\usepackage{tikz}
\usepackage{soul}
\usepackage{hyperref}
\usepackage{nameref}
\usetikzlibrary{quantikz2}
\newtheorem{thm}{Theorem}
\usepackage[normalem]{ulem}

\usepackage{amsthm}

\newtheorem{corollary}{Corollary}

\pdfoutput=1
\hypersetup{colorlinks=true,citecolor=blue,linkcolor=cyan,urlcolor=blue,filecolor= green, breaklinks=true}
\usepackage{url}
\usepackage{breakurl}
\makeatother

\usepackage{mathrsfs}

\begin{document}

\author{Henrique Guerra\href{https://orcid.org/0000-0002-6272-9054}{\includegraphics[scale=0.05]{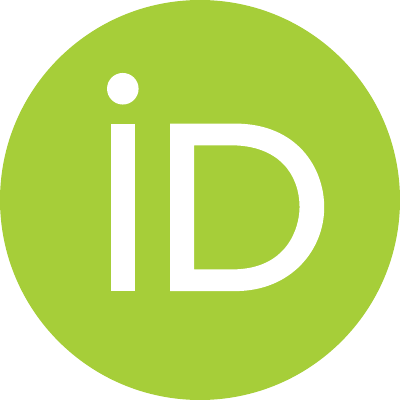}}}
\email{hvsguerra@usp.br}
\affiliation{University of S\~{a}o Paulo, 05508-090, S\~{a}o Paulo,
Brazil}

\author{Tailan S. Sarubi\href{https://orcid.org/0009-0009-2046-6346}{\includegraphics[scale=0.05]{orcidid.pdf}}}
\email{sarubi.santos@gmail.com}
\affiliation{Physics Department, Federal University of Rio Grande do Norte, Natal, 59072-970, Rio Grande do Norte, Brazil}
\affiliation{International Institute of Physics, Federal University of Rio Grande do Norte, 59078-970, Natal, Brazil}

\author{Rafael Chaves\href{https://orcid.org/0000-0001-8493-4019}{\includegraphics[scale=0.05]{orcidid.pdf}}}
\email{rafael.chaves@ufrn.br}
\affiliation{International Institute of Physics, Federal University of Rio Grande do Norte, 59078-970, Natal, Brazil}
\affiliation{School of Science and Technology, Federal University of Rio Grande do Norte, Natal, Brazil}

\author{Jonas Maziero\href{https://orcid.org/0000-0002-2872-986X}{\includegraphics[scale=0.05]{orcidid.pdf}}}
\email{jonas.maziero@ufsm.br}
\affiliation{Physics Department, 
Federal University of Santa Maria, 97105-900,
Santa Maria, RS, Brazil}

\selectlanguage{english}

\title{
Entanglement distribution in quantum networks \\
via swapping of partially entangled pure states 
}

\begin{abstract}
The entanglement swapping protocol (ESP) is a fundamental primitive for distributing quantum correlations across distant nodes in a quantum network. Recent studies have demonstrated that even when the involved qubit pairs are only partially entangled, it is still possible to concentrate and transmit entanglement via Bell-basis measurements. In this work, we extend these ideas to quantum networks with various topologies--including linear, star, and hybrid configurations--by analysing the application of the ESP to initially partially entangled pure states. We investigate how entanglement evolves under such protocols by considering the transformations of the initial states and evaluating the success probabilities for generating maximally entangled states at the output. Our results offer new insights into the dynamics of the entanglement distribution in quantum networks.
\end{abstract}


\date{\today}

\maketitle

\section{Introduction}
\label{intro}

Entanglement is one of the most profound and fundamental features of quantum systems, arising from the inseparability of composite states and the presence of correlations that cannot, in general, be explained by classical local theories. Since the seminal discussions surrounding the Einstein-Podolsky-Rosen (EPR) paradox \cite{PhysRev.47.777} and the formulation of Bell’s theorem \cite{PhysicsPhysiqueFizika.1.195}, it has become evident that entanglement challenges classical notions of locality and realism, reshaping our understanding of physical reality. Beyond its foundational significance, entanglement has also emerged as a key resource for quantum technologies, enabling advances in quantum cryptography \cite{Pirandola}, quantum metrology \cite{Giovannetti}, and quantum computing \cite{Ladd}.

Beyond its foundational role in quantum theory, entanglement serves as a fundamental resource for quantum information processing \cite{nielsen2010quantum}, enabling tasks such as quantum teleportation \cite{Bennett1993}, secure key distribution \cite{Bennett2014}, and distributed quantum computation \cite{Caleffi}. Both early and recent studies—including investigations of entanglement distribution in pure-state networks and proposals for quantum teleportation and quantum repeaters \cite{Azuma2023, Perseguers2008}—have demonstrated that even partially entangled states can be effectively exploited for practical implementations. Through entanglement concentration and distillation protocols, such nonmaximal correlations can be probabilistically converted into high-fidelity entangled states suitable for realistic applications \cite{Bennett1996, Ecker2021, Kraft2021, Cong2025, Koniorczyk2005, Shi2000}.

In this context, quantum networks naturally arise as the fundamental architecture for the large-scale distribution and manipulation of entanglement \cite{Acn2007}. These networks consist of spatially separated nodes interconnected by entangled states, as envisioned in both theoretical models and experimental implementations \cite{Acn2007, Carvacho,Poderini,Zangi2023}. By enabling long-distance quantum communication, quantum networks not only broaden the scope of quantum information applications, but also represent a crucial step toward the realization of the long-anticipated Quantum Internet \cite{Wei2022,Wehner,Kimble,Brito}.

The entanglement swapping protocol (ESP) \cite{Zukowski,Pan} plays a fundamental role in quantum networks \cite{Acn2007}, as it enables the entanglement of distant qubits through operations performed on intermediate entangled pairs \cite{Bergou2021, Maziero, Basso}. By effectively extending the entanglement beyond the range of direct interactions, this protocol facilitates the distribution of quantum correlations across long distances--an essential capability for the development of large-scale quantum communication infrastructures \cite{Mastriani2023, Salimian2023,Memmen}.

Historically, entanglement swapping was introduced by \.{Z}ukowski \textit{et al.} as the foundational mechanism enabling entanglement between distant systems via a Bell-basis measurement on intermediate pairs \cite{Zukowski}. Subsequent work by Bose, Vedral, and Knight showed that variants of entanglement swapping protocol can purify partially entangled states and identified a conserved entanglement quantity in that process \cite{Bose1998}. From a state-discrimination perspective, Sol\'s-Prosser \textit{et al.} analyzed deterministic and probabilistic swapping of nonmaximally entangled states, linking standard ESP to optimal minimum-error strategies and showing probabilistic improvements via maximum-confidence discrimination \cite{SolisProsser2014}. More recently, Bergou \textit{et al.} developed average-concurrence propagation rules across swapping chains \cite{Bergou2021}, while Cong and Xu established formal bounds for the optimal output concurrence achievable via swapping \cite{Cong2025}. In contrast to these lines, our emphasis is not on ensemble-averaged entanglement or optimal bounds, but on closed-form, non-asymptotic success probabilities to herald maximally entangled (Bell) outcomes in chain, star, and hybrid network motifs starting from identical partially entangled inputs. Our analysis yields the full output ensemble and its binomial weight structure, providing operational design quantities (e.g., expected number of swap stages/post-selections) directly relevant to heralded, post-selected distribution.

In this article, we investigate the effects of entanglement swapping applied to partially entangled pure states in various quantum network architectures. In Section \ref{sec:linear}, we begin with linear networks, analysing how successive entanglement swappings on partially entangled pairs can still yield states with a high degree of entanglement. In Section \ref{sec:star}, we turn to star-shaped networks \cite{Cavalcanti,Tavakoli,Andreoli,Poderini}, where a central node shares entangled pairs with all other nodes and, through local operations and measurements, enables the generation of GHZ-type states. Section \ref{sec:starlinear} explores hybrid architectures that combine linear and star-shaped topologies, examining how the resources of both can be integrated. Finally, in Section \ref{sec:FR}, we summarize our findings and outline potential directions for future research.

\section{Entanglement swapping in quantum networks}
\label{sec:ES_qnets}

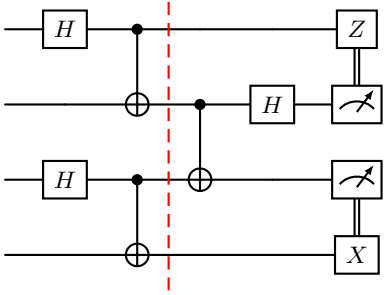
\begin{figure}[t]
    \centering
\begin{quantikz}
    & \gate{H} & \ctrl{1} \slice{} & & & \gate{Z} \\
    && \targ{} & \ctrl{1} & \gate{H} & \meter{} \vcw{-1}\\
    & \gate{H} & \ctrl{1} & \targ{} & & \meter{} \vcw{1}\\
    && \targ{} & & & \gate{X}
\end{quantikz}
    \caption{
Schematic of the entanglement swapping protocol. The segment before the red dashed line represents the initial preparation of two entangled qubit pairs. The segment after the line illustrates the entanglement swap operation, performed by measuring one qubit from each pair in an entangled basis--specifically, the Bell basis in this case. After that, classical information is sent to the end nodes, that perform local operations to prepare a fixed maximally entangled state between systems that have never interacted directly.
}
    \label{circ}
\end{figure}

An entanglement swapping protocol \cite{Zukowski} is a sequence of operations that (in its simplest form) takes two entangled qubit pairs and transforms them into a new entangled pair between previously uncorrelated qubits. This protocol plays a crucial role in overcoming distance-related limitations in entanglement distribution \cite{Bose1998}. Since information losses increase exponentially with the length of the optical fiber \cite{Gisin2015} and thus with distance, intermediate nodes are required to establish entanglement between distant parties who wish to communicate. The prototype entanglement swapping protocol works as illustrated with the quantum circuit in Fig. \ref{circ}.

In the circuit diagram, the first line represents the qubit held by node A, the two middle lines correspond to the qubits of node B, and the fourth line represents the qubit of node C. The protocol begins by entangling A's qubit with one of B's qubits, and the other qubit of B with the qubit of C. The entanglement swap is then performed by applying a Bell-basis measurement to the two qubits held by B. Notably, once the initial entanglement is established, only B's qubits need to interact locally for the protocol to generate entanglement between the distant nodes A and C.

While the entanglement swapping illustrated above is instructive, it involves pairs of qubits that are initially maximally entangled. In contrast, our objective is to investigate how the protocol performs when applied to partially entangled states. We explore this question within the context of different network topologies, including linear and star-shaped configurations, as well as hybrid structures that combine elements of both.

\subsection{Linear networks}
\label{sec:linear}

\begin{figure}[t]
    \centering
    \includegraphics[width=.9\linewidth]{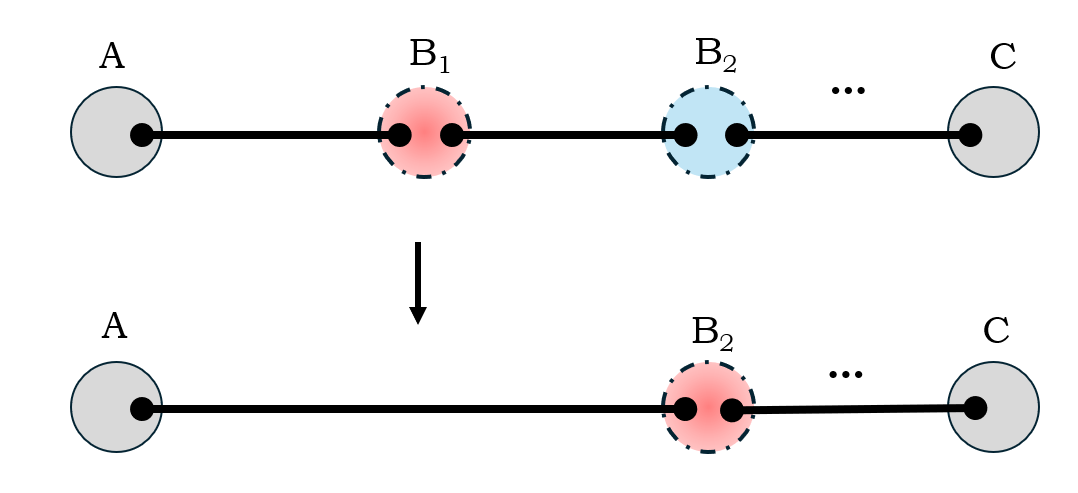}
    \caption{Entanglement exchange process in a linear network. The dots represent qubits, and the qubits connected by edges are entangled. $A$, $B_1$, $B_2$, and $C$ represent nodes in the network, and we are interested in achieving quantum communication between nodes $A$ and $C$, which is enabled by sharing entangled qubits. Through successive entanglement swappings, the intermediate nodes $B_i$ enable nodes $A$ and $C$ to establish entanglement even though they have never interacted directly.}
    \label{eswap}
\end{figure}

In this section, we analyze the propagation of entanglement through a linear network composed of successive entanglement swapping operations, as illustrated in Fig. \ref{eswap}. Each intermediate node performs a swap between pairs of partially entangled states, enabling the establishment of entanglement between the end nodes of the network \cite{Perseguers2008}.

To formalize our analysis, we define a four-dimensional Hilbert space $H$ containing the bipartite entangled states of interest. We begin by considering a partially entangled two-qubit state, given by
$\ket{\eta} = a\ket{00} + b\ket{11}$ with $a,b\in\mathbb{C}$  and
$|a|^2+|b|^2=1$.
We then define
\begin{equation}
    \mathcal{H} = \left\{ \ket{\eta_m}=\frac{a^m\, \ket{00} + b^m\, \ket{11}}{n_m} \text{ for } m \in \mathbb{N} \right\},
    \label{etan}
\end{equation}
with normalization factor being
$n_m = \sqrt{|a|^{2m} + |b|^{2m}}$. The restrictions for $a$ and $b$ yield $n_1=1$.

We write a two-qubit Schmidt form as $\ket{\psi}=a\,\ket{00}+b\,e^{i\phi}\ket{11}$ with \(a,b\in\mathbb{C}\) and \(|a|^2+|b|^2=1\). Global phases are irrelevant, and the relative phase \(e^{i\phi}\) can be absorbed by local phase rotations (e.g., single-qubit \(Z\) rotations), so all probabilities and success rates depend only on \(|a|\) and \(|b|\).
Whenever superpositions of outcomes with different ``exponent-index'' classes are formed, we include the normalization factors \(n_m\) explicitly and track them consistently throughout the derivations.

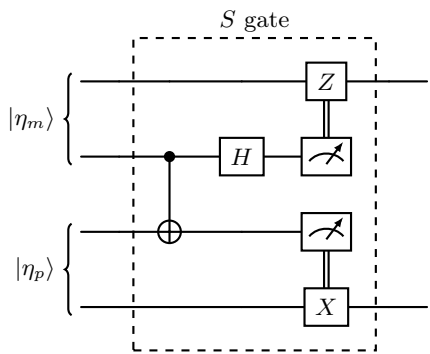
\begin{figure}[t]
    \centering
\begin{quantikz}
    \lstick[2]{$\ket{\eta_m}$}& &\gategroup[4,steps=3, style={dashed,inner sep=6pt}]{$S$ map} & & \gate{Z} & &\\
    & & \ctrl{1} & \gate{H} & \meter{} \vcw{-1}\\
    \lstick[2]{$\ket{\eta_p}$}& & \targ{} & & \meter{} \vcw{1}\\
    && & & \gate{X} & &
\end{quantikz}
    \caption{
Schematic of the entanglement swapping protocol between two partially entangled $\ket{\eta_m},\ket{\eta_p}$ states, outputting an entangled state which is either $\ket{\eta_{m+p}}$ or $\ket{\eta_{m-p}}$. This procedure is what we will from now on refer to as the $S$ map, that acts upon four qubits and outputs two qubits. Whether the output is $\ket{\eta_{m+p}}$ or $\ket{\eta_{m-p}}$ can be determined by the measurements of second and third qubits. As we are not interested in what is the final state of a specific entanglement swap iteration, but rather in the possible states generated by it and its probabilities, we will not use that information.}
    \label{S}
\end{figure}

We will now define the result of an entanglement swap over two states in the set $\mathcal{H}$. To do that, we will define a map $S$ that performs an entanglement swap between two-qubit states $\ket{\eta_m}$ and $\ket{\eta_p}$, with $m \geq p$. The circuit is as shown in Fig. \ref{S}. 

The global state of the system can be described as
\begin{align}
&\ket\psi=\ket{\eta_m}\otimes\ket{\eta_p} \nonumber \\
&=\frac{1}{n_mn_p}\big(a^{m+p}\ket{0000}+a^mb^p\ket{0011}+a^pb^m\ket{1100}\nonumber \\
& \hspace{1.5cm} + b^{m+p}\ket{1111}\big).
\end{align}
As shown in Fig. \ref{S}, the protocol consists of a $CNOT$ gate applied to the third qubit with the second qubit as a control, followed by a Hadamard gate $H$ applied to the second qubit. After applying the gates, we can rearrange these terms as follows
\begin{equation}
\begin{aligned}
    \ket\phi=\frac{1}{n_mn_p\sqrt2}[(a^{m+p}\ket{0000}+b^{m+p}\ket{1001})\\
    +(a^{m+p}\ket{0100}-b^{m+p}\ket{1101})]\\
    +\frac{1}{n_mn_p\sqrt2}[(a^mb^p\ket{0011}+a^pb^m\ket{1010}) \\
    + (a^mb^p\ket{0111}-a^pb^m\ket{1110})].
\label{5}
\end{aligned}
\end{equation}

After the measurement of the second and third qubits, the $Z$ and $X$ gates, which are controlled by classical information,
are applied. The possible outputs of the $S$ gate are either
\begin{align}
& \frac{a^{m+p}\ket{00} + b^{m+p}\ket{11}}{\sqrt{|a|^{2(m+p)}+|b|^{2(m+p)}}} \\
& \text{or} \nonumber \\
& \frac{a^mb^p\ket{00} + a^pb^m\ket{11}}{\sqrt{|a|^{2m}|b|^{2p}+|a|^{2p}|b|^{2m}}} = \frac{a^mb^p\ket{00} + a^pb^m\ket{11}}{|a|^p|b|^p\sqrt{|a|^{2(m-p)}+|b|^{2(m-p)}}} \\ 
& =\frac{a^{m-p}\ket{00} + b^{m-p}\ket{11}}{\sqrt{|a|^{2(m-p)}+|b|^{2(m-p)}}}\text{, minus a global phase.} \nonumber 
\end{align}
That is, we get either $\ket{\eta_{m+p}}$ or $\ket{\eta_{m-p}}$. It follows from Eq. (\ref{5}) that the probability of obtaining  the state $ \ket{\eta_{m+p}}$ is
\begin{equation}
    P(\eta_{m+p}) = \frac{|a|^{2(m+p)}+|b|^{2(m+p)}}{n_m^2n_p^2}=\frac{n_{m+p}^2}{n_m^2n_p^2},
\end{equation}
while the complementary outcome occurs with probability
\begin{equation}
    P(\eta_{m-p}) = \frac{|a|^{2m}|b|^{2p}+|a|^{2p}|b|^{2m}}{n_m^2n_p^2}=|ab|^{2p}\frac{n_{m-p}^2}{n_m^2n_p^2},
\end{equation} 
with $P(\eta_{m+p})+P(\eta_{m-p})=1.$

The entanglement swap map (ESM), that is called here 
as the \(S\) procedure, maps two pairs of two-qubit states to one two-qubit state. 
The initial state $|\eta_m\rangle\otimes|\eta_p\rangle$, under the action of \(S\), 
evolves to
\begin{eqnarray}
\label{troca}
S(\ket{\eta_m}\otimes \ket{\eta_p})=
 \begin{cases}
\ket{\eta_{m+p}},\ \text{with prob. } \frac{n_{m+p}^2}{n_m^2},\\
\ket{\eta_{m-p}},\ \text{with prob. } |ab|^2\frac{n_{m-p}^2}{n_m^2}.
\end{cases}
\end{eqnarray}

Let us consider the case in which $m=p=1$. Since $n_1=1$, the results for the swap are as follows:
\begin{eqnarray}
 S(\ket{\eta} \otimes \ket{\eta}) = 
 \begin{cases}
\ket{\eta_2},\ \text{with prob. } n_2^2=|a|^4+|b|^4, \\
\ket{\eta_{0}},\ \text{with prob. } |ab|^2 n_0^2=2|ab|^2,
\end{cases}
\end{eqnarray}
where we use the notation $|\eta\rangle=|\eta_1\rangle$.

In the case $p=1$, with arbitrary  $m$, we have 
\begin{eqnarray}
 S(\ket{\eta_m} \otimes \ket{\eta}) = 
 \begin{cases}
\ket{\eta_{m+1}},\ \text{with prob. } \frac{n_{m+1}^2}{n_m^2}, \\
\ket{\eta_{m-1}},\ \text{with prob. } |ab|^2\frac{n_{m-1}^2}{n_m^2}.
\end{cases}
\end{eqnarray}

Finally, substituting $m=1,p=0$ into Eq. (\ref{troca}) yields 
\begin{equation}
    S(\ket{\eta_0}\otimes \ket\eta)=\ket\eta
    \label{0}
\end{equation}

Now, consider a linear network with multiple $\ket{\eta}$ states, of which we are interested in obtaining the final state after a series of entanglement swaps.
We proceed to apply entanglement swap operations sequentially between pairs of $\ket\eta$ states. After the first ESM, the output state will be 
\begin{eqnarray}
\ket{A_1} =
\begin{cases}
\ket{\eta_2},\ \text{with prob. } n_2^2=|a|^4+|b|^4, \\
\ket{\eta_{0}},\ \text{with prob. } |ab|^2 n_0^2=2|ab|^2.
\end{cases}
\end{eqnarray}

After the second ESM, we have
\begin{eqnarray}
\ket{A_2} =
 \begin{cases}
\ket{\eta_3},\ \text{with prob. } n_2^2\frac{n_3^2}{n_2^2}=|a|^6+|b|^6, \\
\ket{\eta_1},\ \text{with prob. } n_2^2|ab|^2\frac{n_1^2}{n_2^2}=|ab|^2 \\
\ket{\eta_{1}},\ \text{with prob. } |ab|^2 n_0^2=2|ab|^2.
\end{cases}
\end{eqnarray}
The first two lines of the last equation accrue from the $\ket{\eta_3}$ ket in $\ket{A_1}$, and the last line from the $\ket{\eta_0}$ ket. Because $n_1^2=1\Rightarrow 3|ab|^2=3|a|^4|b|^2+3|a|^2|b|^4$, we can more instructively rewrite $\ket{A_2}$ as
\begin{eqnarray}
 \begin{cases}
\ket{\eta_3},\ \text{with prob. } n_3^2=|a|^6+|b|^6, \\
\ket{\eta_1},\ \text{with prob. } 3|ab|^2n_1^2=3|a|^4|b|^2+3|a|^2|b|^4.
\end{cases}
\end{eqnarray}
This suggests the occurrence of a binomial distribution for $A_n$. Theorems 1, 2 and Corollary 1 prove this assertion.

\begin{thm}
Let $\ket {A}=\ket \eta ^{\otimes (x+1)}$ and $S^{\otimes (x-1)}\ket A = \ket{A_{x-1}} \otimes \ket {\eta}$. If $x-1$ is even and $\ket{A_{x-1}}$ is given by

\begin{eqnarray} 
 \begin{cases}
\ket{\eta_x},\ \text{with prob. } |a|^{2x}+|b|^{2x},\\
\ket{\eta_{x-2}},\ \text{with prob. } x|a|^{2x-2}|b|^2+x|a|^2|b|^{2x-2},\\
\vdots\\
\ket{\eta_{1}},\ \text{with prob. } \binom{x}{\frac{x+1}{2}}|a|^{x+1}|b|^{x-1}+\\ + \binom{x}{\frac{x-1}{2}}|a|^{x-1}|b|^{x+1},
\end{cases}
\end{eqnarray}
then $\ket{A_{x}} $ is equal to
\begin{eqnarray}
\begin{cases}
\ket{\eta_{x+1}},\ \text{with prob. } n^2_{x+1},\\
\ket{\eta_{x-1}},\ prob  (x+1)|a|^{2(x+1)-2}|b|^2 + \\ + (x+1)|a|^2|b|^{2(x+1)-2}, \\
\vdots \\
\ket{\eta_{0}},\ \text{with prob. } \binom{x+1}{\frac{x+1}2}|a|^{x+1}|b|^{x+1}.
\end{cases}
\end{eqnarray}

\end{thm}

\begin{proof}
We have that
$$S^{\otimes(x-1)}\ket A=\ket{A_{x-1}}\otimes\ket \eta$$
Since $x$ is odd, we can rewrite $\ket{A_{x-1}}$ as
\begin{eqnarray}
\ket{A_{x-1}} =
 \begin{cases}
\ket{\eta_{x-2n}},\ \text{with prob. }  \\
\binom{x}{n}|a|^{2x-2n}|b|^{2n}+\binom{x}{x-n}|a|^{2n}|b|^{2x-2n},\\
0 \leq n \leq \frac{x-1}2
\end{cases}
\end{eqnarray}

The $x$-th swap yields
\begin{eqnarray}
 &S^{\otimes x}\ket A =S\ket{A_{x-1}} = \\ \nonumber
 &\begin{cases}
\ket{\eta_{x-2n+1}},\ \text{with prob. }  \\
(\binom{x}{n}|a|^{2x-2n}|b|^{2n}+\binom{x}{x-n}|a|^{2n}|b|^{2x-2n})\frac{n^2_{x-2n+1}}{n^2_{x-2n}},\\
\ket{\eta_{x-2n-1}},\ \text{with prob. }  \\
(\binom{x}{n}|a|^{2x-2n}|b|^{2n}+\binom{x}{x-n}|a|^{2n}|b|^{2x-2n})|ab|^2\frac{n^2_{x-2n-1}}{n^2_{x-2n}},\\
0 \leq n \leq \frac{x-1}2.
\end{cases}
\end{eqnarray}

We can rewrite that as follows
\begin{eqnarray}
\ket{A_{x}} =
 \begin{cases}
\ket{\eta_{x-2n+1}},\ \text{with prob. }  \\
\binom{x}{n}|a|^{2n}|b|^{2n}n^2_{x-2n+1},\\
\ket{\eta_{x-2n-1}},\ \text{with prob. }  \\
\binom{x}{n}|a|^{2n+2}|b|^{2n+2}n^2_{x-2n-1},\\
0 \leq n \leq \frac{x-1}2.
\end{cases}
\end{eqnarray}

Note that, when $ 0<n<\frac{x-1}2$, both $n$ and $n-1$ produce a $\ket{\eta_{x-2n+1}}$ term, because $\ket{\eta_{x-2(n-1)-1}}=\ket{\eta_{x-2n+1}}$; the same applies for $n$ and $n+1$ regarding $\ket{\eta_{x-2n-1}}$. Therefore, we can rewrite $\ket{A_x}$ as
\begin{eqnarray}
\ket{A_{x}} =
 \begin{cases}
\ket{\eta_{x+1}},\ \text{with prob. } n^2_{x+1}\\
\ket{\eta_{0}},\ \text{with prob. } \binom{x}{\frac{x-1}2}|a|^{x+1}|b|^{x+1}n^2_{0}\\
\ket{\eta_{x-2n+1}},\ \text{with prob. }  \\
\left( \binom{x}{n}+\binom{x}{n-1}\right)|a|^{2n}|b|^{2n}n^2_{x-2n+1},\\
0 < n \leq \frac{x-1}2
\end{cases}
\end{eqnarray}

And because $2\binom{x}{\frac{x-1}2}=\binom{x+1}{\frac{x+1}2}$ and $\binom{x}{n}+\binom{x}{n-1}=\binom{x+1}{n}$:
\begin{eqnarray}
\ket{A_{x}} =
 \begin{cases}
\ket{\eta_{x+1}},\ \text{with prob. } n^2_{x+1}\\
\ket{\eta_{0}},\ \text{with prob. } \binom{x+1}{\frac{x+1}2}|a|^{x+1}|b|^{x+1}\\
\ket{\eta_{x-2n+1}},\ \text{with prob. }  \\
 \binom{x+1}{n}|a|^{2n}|b|^{2n}n^2_{x-2n+1},\\
0 < n \leq \frac{x-1}2,
\end{cases}
\end{eqnarray}
we shall have
\begin{eqnarray}
\ket{A_{x}} =
 \begin{cases}
\ket{\eta_{x+1-2n}},\ \text{with prob. }  \\
 \binom{x+1}{n}|a|^{2n}|b|^{2n}n^2_{x+1-2n},\\
0 \leq n \leq \frac{x+1}2
\end{cases}
\label{x case}
\end{eqnarray}
\end{proof}

\begin{thm}
Let $\ket A = \ket \eta ^{\otimes(x+2)}$ and $S^{\otimes x}\ket A= \ket{A_{x}}\otimes \ket \eta$. If $x$ is odd and
\begin{equation}
\ket{A_x} =
\begin{cases}
\ket{\eta_{x+1}},\ \text{with prob. } n_{x+1}^2, \\[4pt]

\ket{\eta_{x-1}},\ \text{with prob. }
\begin{array}{l}
(x+1)\lvert a\rvert^{2(x+1)-2}\lvert b\rvert^2 \\
+ (x+1)\lvert a\rvert^2\lvert b\rvert^{2(x+1)-2},
\end{array} \\[4pt]

\vdots \\[4pt]

\ket{\eta_{0}},\ \text{with prob. }
\binom{x+1}{\frac{x+1}{2}}\lvert a\rvert^{x+1}\lvert b\rvert^{x+1},
\end{cases}
\end{equation}
then
\begin{equation}
\ket{A_{x+1}} =
\begin{cases}
\ket{\eta_{x+2}},\ \text{with prob. }
\lvert a\rvert^{2(x+2)} + \lvert b\rvert^{2(x+2)}, \\[4pt]

\ket{\eta_{x}},\ \text{with prob. }
\begin{array}{l}
(x+2)\lvert a\rvert^{2(x+2)-2}\lvert b\rvert^2 \\
+ (x+2)\lvert a\rvert^2\lvert b\rvert^{2(x+2)-2},
\end{array} \\[4pt]

\vdots \\[4pt]

\ket{\eta_{1}},\ \text{with prob. }
\begin{array}{l}
\binom{x+2}{\frac{x+3}{2}}
\lvert a\rvert^{x+3}\lvert b\rvert^{x+1} \\
+ \binom{x+2}{\frac{x+1}{2}}
\lvert a\rvert^{x+1}\lvert b\rvert^{x+3}.
\end{array}
\end{cases}
\end{equation}

\end{thm}
\begin{proof}

{Recalling Eq. (\ref{x case})}:
\begin{eqnarray*}
{\ket{A_{x}} =}
\begin{cases}
{\ket{\eta_{x+1-2n}},\ \text{with prob. } } \\
{ \binom{x+1}{n}|a|^{2n}|b|^{2n}n^2_{x+1-2n},}\\
{0 \leq n \leq \frac{x+1}2}.
\end{cases}
\end{eqnarray*}

{The $x+1$-th swap yields}

\begin{eqnarray}
{ S^{\otimes (x+1)}\ket A =S\ket{A_{x}} =}\nonumber\\
 \begin{cases}
{\ket{\eta_{x+1-2n+1}},\ \text{with prob. } } \\
{\binom{x+1}{n}|a|^{2n}|b|^{2n}n^2_{x+1-2n}\frac{n^2_{x+1-2n+1}}{n^2_{x+1-2n}},}\\
{\ket{\eta_{x+1-2n-1}},\ \text{with prob. } } \\
{( \binom{x+1}{n}|a|^{2n}|b|^{2n}n^2_{x+1-2n}|ab|^2\frac{n^2_{x+1-2n-1}}{n^2_{x+1-2n}},}\\
{0 \leq n < \frac{x+1}2,}\\
{\ket{\eta_1}, \text{with prob. } \binom{x+1}{\frac{x+1}2}|a|^{x+1}|b|^{x+1}},
\end{cases}
\end{eqnarray}
where we compute $S\ket{\eta_0}=\ket{\eta_1}$ separately. Thus
\begin{eqnarray}
{\ket{A_{x+1}} =}
 \begin{cases}
{\ket{\eta_{x+2-2n}},\ \text{with prob. } } \\
{\binom{x+1}{n}|a|^{2n}|b|^{2n}n^2_{x+2-2n},}\\
{\ket{\eta_{x-2n}},\ \text{with prob. } } \\
{( \binom{x+1}{n}|a|^{2n+2}|b|^{2n+2}n^2_{x-2n},}\\
{0 \leq n < \frac{x+1}2;}\\
{\ket{\eta_1}, \text{with prob. } \binom{x+1}{\frac{x+1}2}|a|^{x+1}|b|^{x+1}}
\end{cases}
\end{eqnarray}
Rearranging the sum, we get $\ket{A_{x+1}}$:
\begin{eqnarray}
 \begin{cases}
 {\ket{\eta_{x+2}}, \ \text{with prob. } n^2_{x+2},}\\
{\ket{\eta_{x+2-2n}},\ \text{with prob. } } \\
{\left( \binom{x+1}{n}+\binom{x+1}{n-1}\right)|a|^{2n}|b|^{2n}n^2_{x+2-2n},}\\
{0 < n < \frac{x+1}2;}\\
{\ket{\eta_1}, \text{with prob. } \left(\binom{x+1}{\frac{x+1}2}+\binom{x+1}{\frac{x-1}2}\right)|a|^{x+1}|b|^{x+1}}
\end{cases}
\end{eqnarray}

Again, because $\binom{x}{n}+\binom{x}{n-1}=\binom{x+1}{n}$, this last equation equals to
\begin{eqnarray}
{\ket{A_{x+1}} =}
\begin{cases}
{\ket{\eta_{x+2}}, \ \text{with prob. } n^2_{x+2},}\\
{\ket{\eta_{x+2-2n}},\ \text{with prob. } } \\
{ \binom{x+2}{n}|a|^{2n}|b|^{2n}n^2_{x+2-2n},}\\
{0 < n < \frac{x+1}2;}\\
{\ket{\eta_1}, \text{with prob. } \binom{x+2}{\frac{x+1}2}|a|^{x+1}|b|^{x+1}}
\end{cases}
\end{eqnarray}
{or}

\begin{eqnarray}
{\ket{A_{x+1}} =}
\begin{cases}
{\ket{\eta_{x+2-2n}},\ \text{with prob. } } \\
{ \binom{x+2}{n}|a|^{2n}|b|^{2n}n^2_{x+2-2n},}\\
{0 \leq n \leq \frac{x+1}2.}\\
\end{cases}
\end{eqnarray}
\end{proof}

\begin{corollary}
{$\ket{A_x}$ follows a binomial distribution.}
\end{corollary}

\begin{proof}

{Induction has already been proven in Theorems 1 and 2. Now let us provide the base case for $x=2$. As we have already shown,}
\begin{eqnarray}
{ \ket{A_1} =} 
\begin{cases}
{\ket{\eta_2},\ \text{with prob. } n_2^2=|a|^4+|b|^4,} \\
{\ket{\eta_{0}},\ \text{with prob. } |ab|^2 n_0^2=2|ab|^2.}
\end{cases}
\end{eqnarray}
{and}
\begin{eqnarray*}
{ \ket{A_2} =} 
 \begin{cases}
{\ket{\eta_3},\ \text{with prob. } n_3^2=|a|^6+|b|^6,} \\
{\ket{\eta_1},\ \text{with prob. } 3|ab|^2n_1^2=3|a|^4|b|^2+3|a|^2|b|^4}.
\end{cases}
\end{eqnarray*}

Calculating $S\ket{A_2}$ yields
\begin{eqnarray}
{ \ket{A_3} =} 
 \begin{cases}
{\ket{\eta_4},\ \text{with prob. } n_4^2,} \\
{\ket{\eta_2},\ \text{with prob. } 4n_2^2|ab|^2,} \\
{\ket{\eta_0},\ \text{with prob. } 6|ab|^4.}
\end{cases}
\end{eqnarray}
\end{proof}
Now we give a combinatorial intuition of this result. Each Bell-basis measurement in a swapping step effectively increments or decrements an ``exponent index'' by $\pm 1$.
After $x$ independent swaps, the number of paths with exactly $k$ ``$b$-steps'' is $\binom{x}{k}$, producing amplitudes proportional to $a^{\,x-k}b^{\,k}$.
After normalization, the corresponding probability weights are binomial and scale as $\Pr[k]\ \propto\ \binom{x}{k}\,|a|^{2(x-k)}\,|b|^{2k}\,,$ which explains the binomial structure of the outcome ensemble observed in the algebraic derivation.



\begin{figure}[t]
    \centering
\begin{quantikz}[row sep={0.8cm, between origins}]
    \lstick[4]{$\ket{\xi_m}$}& & & & & &\\
    \setwiretype{n}& \vdots& & & & \vdots & \\
    & &\gategroup[4,steps=3, style={dashed,inner sep=6pt}]{$S$ map} & & \gate{Z} & &\\
    & & \ctrl{1} & \gate{H} & \meter{} \vcw{-1}\\
    \lstick[2]{$\ket{\eta_p}$}& & \targ{} & & \meter{} \vcw{1}\\
    && & & \gate{X} & &
\end{quantikz}
    \caption{
Schematic of the entanglement swapping protocol between two partially entangled states $\ket{\xi_m},\ket{\eta_p}$, outputting an entangled state which is either $\ket{\xi_{m+p}}$ or $\ket{\xi_{m-p}}$. It is clear that only one qubit from $\ket{\xi_m}$ participates in the protocol.}  
    \label{multipartido}
\end{figure}
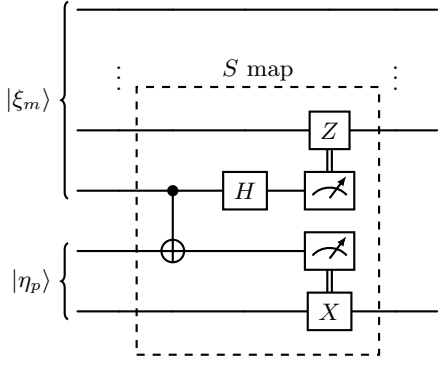

We can extend our results for \textit{multipartite states}. To do that, we can consider the swap between  a $\ket{\eta_m}$-like  multipartite state
\begin{equation}
    \ket{\xi_m}=\frac{a^m\ket{0}^{\otimes j}+b^m\ket{1}^{\otimes j}}{n_m}
    \label{xin}
\end{equation}
and a bipartite  state $\ket{\eta_p}$. The circuit for such a swap is shown in Fig. \ref{multipartido}.

Based on the calculations for \(S(\ket{\eta_m}\otimes\ket{\eta_p})\), we now determine the output of \(S(\ket{\xi_m}\otimes\ket{\eta_p})\). To this end, consider the pure‐state density operator $\rho_i = \ket{\xi_m}\bra{\xi_m}\otimes \ket{\eta_p}\bra{\eta_p} = \ket{\psi}\bra{\psi}$, where $\ket{\psi}$ is:
\begin{equation}
\begin{aligned}
    \ket\psi=\frac{1}{n_mn_p}\big(a^{m+p}\ket{0}^{\otimes j}\ket{00}+a^mb^p\ket{0}^{\otimes j}\ket{11}\\+a^pb^m\ket{1}^{\otimes j}\ket{00}+b^{m+p}\ket{1}^{\otimes j}\ket{11}\big).
\end{aligned}
\end{equation}
Only the qubits at positions $j-1$ and $j$ are involved in the entanglement swap; hence, we express $\ket{\psi}$ as
\begin{equation}
\begin{aligned}
    \ket\psi&=\frac{1}{n_mn_p}\big(a^{m+p}\ket{0}^{\otimes (j-2)}\ket{0000}+a^mb^p\ket{0}^{\otimes (j-2)}\ket{0011}\\ &+a^pb^m\ket{1}^{\otimes (j-2)}\ket{1100}+b^{m+p}\ket{1}^{\otimes (j-2)}\ket{1111}\big).
\end{aligned}
\end{equation}

Therefore, applying the $CNOT$ gate followed by the Hadamard gate transforms the state $\ket\psi$ into one that can be rearranged as 
\begin{equation}
\begin{aligned}
    \frac{1}{\sqrt2 n_mn_p}\big[\big(a^{m+p}\ket{0}^{\otimes (j-2)}\ket{0000}+b^{m+p}\ket{1}^{\otimes (j-2)}\ket{1001}\big)\\
    +\big(a^{m+p}\ket{0}^{\otimes (j-2)}\ket{0100}-b^{m+p}\ket{1}^{\otimes (j-2)}\ket{1101}\big)\\
    +\big(a^mb^p\ket{0}^{\otimes (j-2)}\ket{0011}++a^pb^m\ket{1}^{\otimes (j-2)}\ket{1010}\big)\\
    +\big(a^mb^p\ket{0}^{\otimes (j-2)}\ket{0111}-a^pb^m\ket{1}^{\otimes (j-2)}\ket{1110}\big)\big].
\end{aligned}
\end{equation}
After measurement and the $Z$ and $X$ gates controlled by classical information, the final state will be either
\begin{equation}
\begin{aligned}
\frac{a^{m+p}\ket{0}^{\otimes (j-2)}\ket{00} + b^{m+p}\ket{1}^{\otimes (j-2)}\ket{11}}{\sqrt{|a|^{2(m+p)}+|b|^{2(m+p)}}} = \ket{\xi_{m+p}}
\end{aligned}
\end{equation}
with probability
\begin{equation}P(\xi_{m+p})=\frac{|a|^{2(m+p)}+|b|^{2(m+p)}}{n_m^2n_p^2}=\frac{n_{m+p}^2}{n_m^2n_p^2}
\end{equation}
or
\begin{equation}
\begin{aligned}
\frac{a^mb^p\ket{0}^{\otimes (j-2)}\ket{00} + a^pb^m\ket{1}^{\otimes (j-2)}\ket{11}}{\sqrt{|a|^m|b|^p+|a|^p|b|^m}} = \ket{\xi_{m-p}}
\end{aligned}
\end{equation}
which occurs with probability
\begin{equation}
    P(\xi_{m-p})=\frac{|a|^{2m}|b|^{2p}+|a|^{2p}|b|^{2m}}{n_m^2n_p^2}=|ab|^{2p}\frac{n_{m-p}^2}{n_m^2n_p^2}.
\end{equation} 

In compact form, the entanglement-swap map acts 
as follows:
\begin{eqnarray}
S(\ket{\xi_m}\otimes \ket{\xi_p})=
 \begin{cases}
\ket{\xi_{m+p}},\ \text{with prob. } \frac{n_{m+p}^2}{n_m^2}, \\
\ket{\xi_{m-p}},\ \text{with prob. } |ab|^2\frac{n_{m-p}^2}{n_m^2},
\end{cases}
\end{eqnarray}
which is analogous to Eq. (\ref{troca}), that was obtained for bipartite states. 

The fact that the recurrences obtained are identical means that the theorem derived for linear networks involving bipartite states also applies to networks containing one multipartite state:
\begin{eqnarray}
S^{\otimes x}(\left( \ket \eta^{\otimes x}\otimes \ket \xi\right)=
\begin{cases}
\ket{\xi_{x+1-2n}},\ \text{with prob. }  \\
 \binom{x+1}{n}|a|^{2n}|b|^{2n}n^2_{x+1-2n},\\
0 \leq n \leq \frac{x+1}2
\label{importante}
\end{cases}
\end{eqnarray}

The results of this section establish a systematic framework for linear entanglement–swapping chains, by which successive swap operations propagate and concentrate entanglement across an arbitrary number of qubits, even when beginning with partially entangled initial states. Consequently, these expressions provide the necessary tools to predict end-to-end entanglement distributions in linear topologies. In Sec. \ref{sec:star}, we extend this formalism to star-network architectures, thereby demonstrating its applicability beyond purely linear configurations.

\subsection{Star networks}
\label{sec:star}

\begin{figure}[t]
    \centering
    \includegraphics[width=1\linewidth]{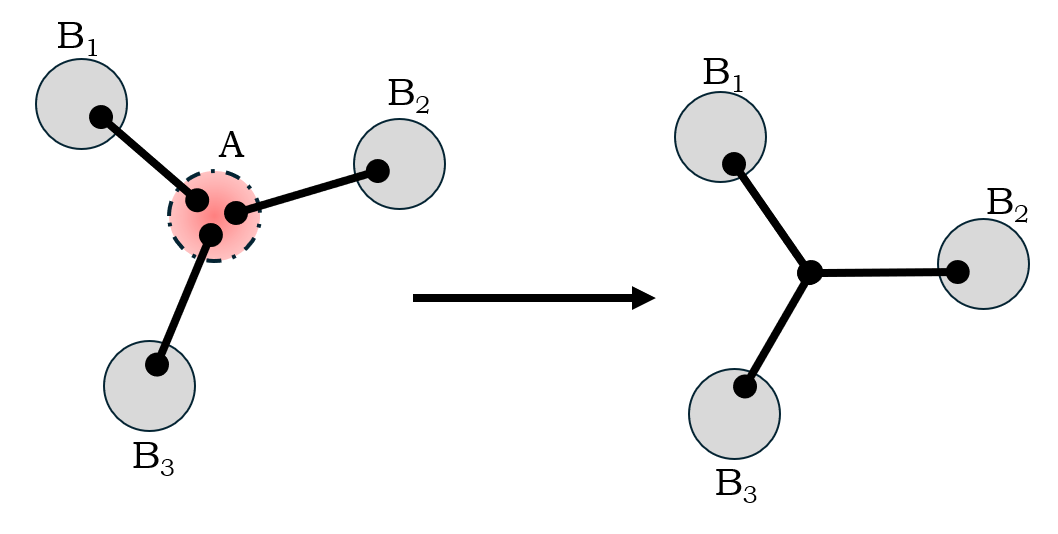}
    \caption{Entanglement swapping process in star-shaped quantum networks. Initially, the central node \( A \) shares entangled qubit pairs with the peripheral nodes \( B_1, B_2, B_3 \). After a joint measurement at \( A \), the entanglement is transferred, establishing direct quantum connections among the peripheral nodes. This process can be generalized to \( B_n \) peripheral nodes.}
    \label{fig:enter-label}
\end{figure}

In star-shaped quantum networks, illustrated in Fig. \ref{fig:enter-label}, a central node \( A \) initially shares \( n \) entangled qubit pairs with peripheral nodes \( B_1, B_2, \ldots, B_n \). Each pair is described by the partially entangled state  
\begin{equation}
\ket{\eta} = a \ket{00} + b \ket{11}.
\end{equation}

The initial global state of the network is the tensor product of these $y$ pairs: 
\begin{equation}
\ket B = \ket{\eta}^{\otimes y}.
\end{equation}
After the swap is carried out, it is expected that all \( B_1, B_2, \ldots, B_n \) share a GHZ-like state.

\begin{figure}[t]
    \centering
\begin{quantikz}
    \lstick[2]{$\ket{\eta}$}& & \gategroup[6,steps=3, style={dashed,inner sep=6pt}]{$S_2$ map} & & \gate{Z} & & \\
    & & \ctrl{4} & \gate{H} & \meter{} \vcw{-1} \\
    \lstick[2]{$\ket{\eta}$}& & & & \gate{X}& & \\
    & & \targ{} &  & \meter{} \vcw{-1}\\
    \lstick[2]{$\ket{\eta}$}& & & & \gate{X} & & \\
    & & \targ{} & & \meter{} \vcw{-1}
\end{quantikz}
    \caption{
Schematic of a entanglement swapping protocol intended to create a GHZ state. }
    \label{star}
\end{figure}
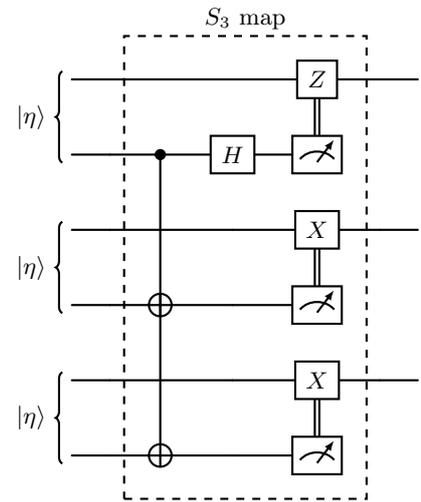

If $y=3$, the circuit that represents the operation we want to carry out is as shown in Fig. \ref{star}. The map defined as $S_{y-1}$ is such that $S_{y-1} \ket B= \ket{B_y}$, which is a $y$-qubit state.

We can demonstrate the general action of $S_y$ applying it to our initial state with arbitrary $y$: 
\begin{equation}
        \begin{aligned}
            &\ket B=\ket\eta^{\otimes y}=\bigotimes_{i=0}^{y}\ket{\eta}=\ket\eta \otimes \left[\bigotimes_{i=0}^{y-1}\ket{\eta}\right]\\
            &=\left(a\ket{00} + b\ket{11}\right) \otimes \left[\bigotimes_{i=0}^{y-1}\left(a\ket{00} + b\ket{11}\right)\right]\\
            &=a\ket{00} \otimes \left[\bigotimes_{i=0}^{y-1}\left(a\ket{00} + b\ket{11}\right)\right]\\&+ b\ket{11} \otimes \left[\bigotimes_{i=0}^{y-1}\left(a\ket{00} + b\ket{11}\right)\right].
        \end{aligned}
\end{equation}

Analogously to the linear case, when performing the swap operation, first applying a CNOT gate and then a Hadamard gate, the resulting state of the process can be expressed as follows:
\begin{equation}
        \begin{aligned}
            &\frac{a\ket{00}}{\sqrt2} \otimes \left[\bigotimes_{i=0}^{y-1}\left(a\ket{00} + b\ket{11}\right)\right]\\&+\frac{b\ket{10}}{\sqrt2}\otimes \left[\bigotimes_{i=0}^{y-1}\left(a\ket{01} + b\ket{10}\right)\right] \\
            &\frac{a\ket{01}}{\sqrt2} \otimes \left[\bigotimes_{i=0}^{y-1}\left(a\ket{00} + b\ket{11}\right)\right]\\&-\frac{b\ket{11}}{\sqrt2} \otimes \left[\bigotimes_{i=0}^{y-1}\left(a\ket{01} + b\ket{10}\right)\right].
        \end{aligned}
\end{equation}
Measuring the second qubit of each term of the product, given a measurement output $\{t_1,\dots, t_y\}, t_i \in \{0,1\}$, and considering
\begin{equation}
    o=\sum\limits_{i=0}^{y}t_i ,
\end{equation}
(i.e., $o$ is equal to the number of outputs $1$), the final state of the system will be
\begin{equation}
\begin{aligned}
\ket{B_y}&=\frac{a^{y-o}b^{o}\ket{t_1\dots t_n}+a^{o}b^{y-o}\ket{(t_1 \oplus 1)\dots (t_n \oplus 1)}}{\sqrt{|a|^{2(y-o)}|b|^{2o}+|a|^{2o}|b|^{2(y-o)}}}\\ &\xrightarrow{cZ\text{, }cX\text{s}}
\frac{a^{y-o}b^{o}\ket{0}^{\otimes y}+a^{o}b^{y-o}\ket{1}^{\otimes y}}{\sqrt{|a|^{2(y-o)}|b|^{2o}+|a|^{2o}|b|^{2(y-o)}}}=\ket{\xi_{|y-2o|}},
\end{aligned}
\end{equation}
if $t_1=0$, and 
\begin{equation}
\begin{aligned}
\ket{B_y}&=\frac{a^{y-o}b^{o}\ket{t_1\dots t_n}-a^{o}b^{y-o}\ket{(t_1 \oplus 1)\dots (t_n \oplus 1)}}{\sqrt{|a|^{2(y-o)}|b|^{2o}+|a|^{2o}|b|^{2(y-o)}}} \\&\xrightarrow{cZ\text{, }cX\text{s}}
\frac{a^{y-o}b^{o}\ket{0}^{\otimes y}+a^{o}b^{y-o}\ket{1}^{\otimes y}}{\sqrt{|a|^{2(y-o)}|b|^{2o}+|a|^{2o}|b|^{2(y-o)}}}=\ket{\xi_{|y-2o|}}.
\end{aligned}
\end{equation}

Therefore, without knowledge of $\{t_1,\dots, t_n\}$, we can represent our final system as:
\begin{eqnarray}
{\ket{B_{y}} =}
\begin{cases}
{\ket{\xi_{y-2o}},\ \text{with prob. } } \\
{ \binom{y}{o}\left(|a|^{2y-2o}|b|^{2o}+|a|^{2o}|b|^{2y-2o}\right)}\\
{0 \leq o \leq \frac{y}2}.
\end{cases}
\end{eqnarray}

Thus
\begin{eqnarray}
    S_{y-1}\ket{\eta}^{\otimes y} =
    \begin{cases}
    {\ket{\xi_{y-2n}},\ \text{with prob. } } \\
{ \binom{y}{n}|a|^{2n}|b|^{2n}n_{y-2n}^2}\\
{0 \leq n \leq \frac{y}2}
    \end{cases}
\label{estrela}
\end{eqnarray}

Recalling Eq. (\ref{importante}), this means that
\begin{equation}
    S_z \ket{\eta}^{\otimes (z+1)}=S^{\otimes z} \left(\ket{\eta}^{\otimes z} \otimes \ket{\xi}  \right) ,
\label{uau!}
\end{equation}
and, more importantly, that 
\begin{equation}
    \begin{aligned}
        &S^{\otimes x}\Big\{ \ket{\eta}^{\otimes x} \otimes \big[ S_y \ket{\eta}^{\otimes (y+1)}   \big] 
         \Big\} \\
        &= S^{\otimes x} \Big[\ket{\eta}^{\otimes x} \otimes \big( S^{\otimes y} \big( \ket{\eta}^{\otimes y} \otimes \ket{\xi}  \big)  \big) \Big] \\
        &= S^{\otimes(x+y)} \Big(\ket{\eta}^{\otimes(x+y)} \otimes \ket{\xi} \Big) .
    \end{aligned}
\label{uau!!}
\end{equation}

A single Bell-basis measurement at the center of the star converts a set
of partially entangled pairs into GHZ-like states
whose 
probabilities are fixed by the initial entanglement, allowing for the single-step generation of high-fidelity multiparty links. Adjusting the preparation coefficients allows us to smoothly balance the probability of success with the quality of the final entanglement to meet different protocols requirements. This unified treatment naturally recovers both the two- and three-node cases, highlighting the intrinsic efficiency and tunability of the star topology for distributing multiparty entanglement. In this way, we combine the efficient distribution of the star with the long-distance reach of sequential entanglement exchanges, providing a versatile model for hybrid quantum networks, which will be discussed in the next section.



\subsection{Star-linear networks}
\label{sec:starlinear}

\begin{figure}[t]
    \centering  \includegraphics[width=1\linewidth]{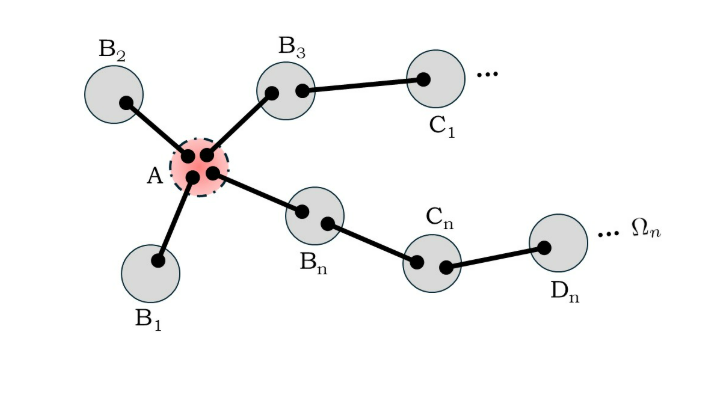}
    \caption{Example of a star-linear quantum network. The central node \(A\) shares entangled pairs with the peripheral nodes \(B_1, \dots, B_n\). One of these nodes, \(B_n\), is further connected to a linear chain of nodes \((C_n, D_n, \dots)\) leading to the final node \(\Omega_n\). To establish a multipartite ``conference'' among all \(\Omega_n\), one first creates a conference among all \(B_n\) (as described in Section~B). Then, by applying the \(*\) operator successively, \(B_n\) is replaced by \(C_n\) in the conference, \(C_n\) is replaced by \(D_n\), and so on, until ultimately reaching \(\Omega_n\).}
    \label{fig:enter-label1}
\end{figure}

The star-linear topology integrates the efficiency of star networks with the scalability of linear chains, enabling robust multipartite entanglement distribution across large-scale quantum networks shown in Fig. \ref{fig:enter-label1}. The protocol proceeds in two stages.

In the star-swap stage, the central node performs joint measurements on $y$ identical pairs $\ket{\eta}^{\otimes y}$, generating a mixture of $y$-partite entangled states among peripheral nodes:
    \begin{equation}
\ket{\xi_{y-2i}} \text{ with prob. } \binom{y}{i} |a|^{2(y-i)} |b|^{2i},
        \label{eq:star_stage}
    \end{equation}
where $i=0,\cdots,y$ and $\ket{\xi_m}$ is defined in Eq. (\ref{xin}).
    
In the linear-Swap stage, for each outcome $i$ from the star-swap stage, peripheral nodes propagate entanglement through $x$ sequential swaps with $\ket{\eta}$ pairs, producing bipartite states between end nodes:
    \begin{equation}
     \ket{\eta_{\Delta_{ij}}} \text{ with prob. } \binom{x}{j} |a|^{2(x-j)} |b|^{2j},   \label{eq:linear_stage}
    \end{equation}
with $\Delta_{ij} = |(y-2i) + (x-2j)|$ and $\ket{\eta_m}$ is defined in Eq. (\ref{etan}).

The global state combines both stages through conditional averaging:
\begin{equation}
\ket{\eta_{\Gamma_{ij}}} \text{with prob. } \binom{y}{i} \binom{x}{j} |a|^{2[(y-i)+(x-j)]} |b|^{2(i+j)},
    \label{eq:double_sum}
\end{equation}
where $i=0,\cdots,x$, $j=0,\cdots,y$, and $\Gamma_{ij} = \Delta_{ij}$. 
Remarkably, this expression
simplifies to the binomial form of a single linear chain with $x+y$ links:
\begin{equation}
\ket{\eta_{x+y-2k}} \text{ with prob. } \binom{x+y}{k} |a|^{2(x+y-k)} |b|^{2k}.
    \label{eq:collapsed_form}
\end{equation}

This unified framework demonstrates that star-linear networks preserve the entanglement distribution properties of linear chains while leveraging multipartite efficiency. The emergent binomial structure warants efficient concentration of entanglement across arbitrary distances, establishing this architecture as a fundamental building block for 
quantum networks. Decoupling central control after the star stage enhances robustness against single-point failures, making the topology ideal for scalable quantum infrastructures.

\section{Final Remarks}
\label{sec:FR}

\begin{figure}[t]
    \centering  \includegraphics[width=1\linewidth]{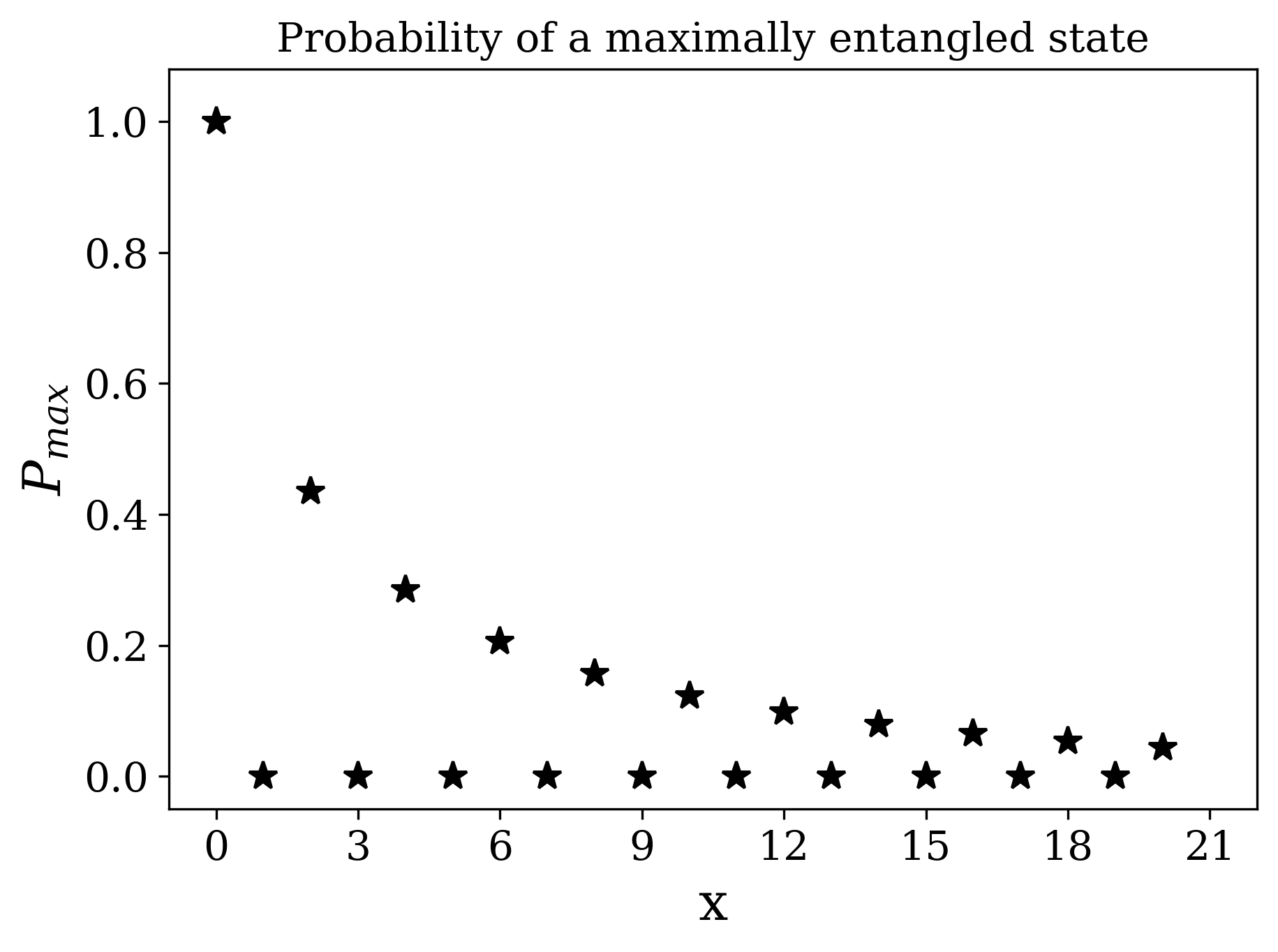}
    \caption{Value of $P_{max}$ as $x$ increases, for $a=0.566$. When $x$ is even, $P_{max}$ follows Eq. \ref{Pmax}; when $x$ is odd, there is no possibility of $\ket{\eta_0}$ occurring, and thus $P_{max}=0$.}
    \label{fig:pmax}
\end{figure}

The results presented in this work show that the probability of obtaining a maximally entangled state after entanglement swaps in a star-linear network is
\begin{equation}
    P_{max}=\binom{x}{\frac{x}2}a^xb^x
    \label{Pmax}
\end{equation}
if $x$ is even. The curve of this function is portrayed in Fig. \ref{fig:pmax}. End parties can identify whether they share a maximally entangled state through the measurements performed by the intermediate parties through each step, which we ignored since our purpose was solely to show the distribution of probabilities for the outcomes.  More importantly, this probability shows us the optimal strategy for post-selection: the expected number of attempts to obtain a maximally entangled state when post-selecting only at the end is $P_{max}^{-1}$. Intuitively, this number increases if instead the strategy is to post-select after every two swaps (that is, whenever possible). Then 
\begin{equation}
    \langle N\rangle =\Big(\frac{1}{2a^2b^2}\Big)^{\frac{x}2}=\frac12a^{-x}b^{-x}.
\end{equation}
This number is not only greater than what we obtained for the post-selecting at the end strategy but also grows binomially faster. Using Stirling's approximation, one can verify that
\begin{equation}
    \binom{x}{\frac{x}2}\approx \frac{2^x}{\sqrt{\frac{x\pi}{2}}},
\end{equation}
meaning the ratio between the expected values for each strategy grows a little less than exponentially. That is a great difference that should be considered for entanglement distribution in networks.

The results presented in this work highlight the central role of entanglement swapping in the design and operation of quantum networks. In particular, we have shown that, within a linear topology, it is possible to obtain a maximally entangled state  from a sequence of partially entangled states. This effectively realizes a form of entanglement purification (or concentration) via consecutive swap operations. 
Moreover, the existence of a nonzero probability \( P \) of achieving a maximally entangled state reinforces the practical viability of key quantum communication protocols, such as teleportation and quantum key distribution, across distributed network architectures.


In the case of star-shaped quantum networks, the key innovation lies in the ability to redistribute quantum correlations through local operations--namely, CNOT and Hadamard gates--followed by measurements on the central node. This process projects the peripheral nodes into GHZ-type entangled states, effectively transferring the initial pairwise entanglement to a genuinely multipartite configuration. The success probability of this transformation depends sensitively on the asymmetry between the entanglement coefficients \( a \) and \( b \), which directly influence the degree and quality of the final correlations. Notably, this mechanism generalizes naturally to systems with more particles, resembling established protocols for multipartite entanglement swapping. Such schemes, historically studied in the context of GHZ-state generation, demonstrate the efficiency of single-node measurements in establishing long-range quantum correlations across a network.


The integration of star and linear network architectures into a unified star-linear topology offers a scalable and modular framework for entanglement distribution. This two-stage protocol--beginning with the generation of a multipartite state among intermediate nodes in a star configuration, followed by its propagation through linear entanglement-swapping chains--enables the construction of complex quantum networks with structured correlation patterns. Crucially, the ability to establish long-range multipartite entanglement without the need to maintain a central control node enhances the robustness and decentralization of the resulting network. This feature is especially valuable for quantum communication infrastructures, where reducing single points of failure is essential for both scalability and security.

Finally, given the growing interest in large-scale quantum networks, the results presented here contribute to the advancement of quantum internet architectures. Future developments will require a deeper understanding of adaptive entanglement strategies, the behavior of mixed states under entanglement-swapping operations, and the extension of these protocols to qudit systems. In particular, exploring the effects of noise on mixed-state entanglement, designing protocols tailored for heterogeneous quantum systems, and investigating alternative network topologies that enhance resource efficiency and scalability are promising directions for further research. These efforts will be essential for building robust, flexible, and globally distributed quantum communication networks.

\begin{acknowledgments}
This work was supported by the Simons Foundation (Grant Number 1023171, RC); the  National Council for Scientific and Technological Development (CNPq) under Grants No. 409673/2022-6 and No. 421792/2022-1, No. 307295/2020-6, No. 403181/2024-0, No. 300083/2025-4; the National Institute for the Science and Technology of Quantum Information (INCT-IQ) under Grant No. 465469/2014-0; Financiadora de Estudos e Projetos (grant 1699/24 IIF-FINEP);  Research Support Foundation of the State of Rio Grande do Sul (FAPERGS) under Grant No. 25/2551-0002608-3; and S\~{a}o Paulo Research Foundation (FAPESP) under Grant No. 19722-0/2024. 
\end{acknowledgments}


\end{document}